\documentclass{iaus}
\usepackage{graphicx}
\usepackage{psfrag}
\usepackage{rotating}
\usepackage{sidecap}
\newlength{\abovecaptionskip}
\newlength{\belowcaptionskip}
\newcommand{\mytextup}[1]{\textup{\scriptsize #1}}

\title[Black hole accretion] 
{Black hole accretion: theoretical limits\break and observational implications}

\author[Heinzeller, D., Duschl, W.J., Mineshige, S. \& Ohsuga, K.]   
{Dominikus Heinzeller$^{1,2}$, Wolfgang J. Duschl$^{1,2,3}$, Shin Mineshige$^4$\break \and Ken Ohsuga$^5$}

\affiliation{$^1$Zentrum f\"{u}r Astronomie Heidelberg, Institut f\"{u}r Theoretische Astrophysik,
Albert-Ueberle-Stra{\ss}e 2, 69120 Heidelberg, Germany\\[\affilskip]
$^2$now at: Institut f\"{u}r Theoretische Physik und Astrophysik, Universit\"{a}t Kiel, 24098 Kiel, Germany;
email: \{hd,wjd\}@astrophysik.uni-kiel.de\\[\affilskip]
$^3$Steward Observatory, The University of Arizona, 933 North Cherry Ave, Tucson,\break AZ 85721, USA\\[\affilskip]
$^4$Yukawa Institute for Theoretical Physics, Kyoto University, Kitashirakawa-Oiwakecho,
Sakyo-ku, Kyoto 606-8502, Japan; email: minesige@yukawa.kyoto-u.ac.jp\\[\affilskip]
$^5$Department of Physics, Rikkyo University, 3-34-1 Nishi-Ikebukuro, Toshimaku,\break Tokyo 171-8501,
Japan; email: k\_ohsuga@rikkyo.ac.jp}

\pubyear{2006}
\volume{238}  
\pagerange{001--999}
\date{??? and in revised form ???}
\jname{Black Holes: from Stars to Galaxies -- across the Range of Masses}
\editors{V. Karas \& G. Matt, eds.}
\begin{document}

\maketitle

\begin{abstract} Recently, the issue of the role of the Eddington limit in
accretion discs became a matter of debate. While the classical (spherical)
Eddington limit is certainly an over-simplification, it is not really clear
how to treat it in a flattened structure like an accretion disc. We
calculate the critical accretion rates and resulting disc
luminosities for various disc models corresponding to the classical Eddington
limit by equating the attractive and repulsive forces locally. We also discuss
the observational appearance of such highly accreting systems by analyzing
their spectral energy distributions. Our calculations indicate that the allowed mass accretion rates
differ considerably from what one expects by applying the Eddington limit in
its classical form, while the luminosities only weakly exceed their classical
equivalent. Depending on the orientation of the disc relative to the observer,
mild relativistic beaming turns out to have an important influence on
the disc spectra. Thus, possible super-Eddington accretion, combined with mild
relativistic beaming, supports the idea that ultraluminous X-ray sources
host stellar mass black holes and accounts partially for the observed
high temperatures of these objects.
\end{abstract}
\firstsection 
\section{Introduction}
With the constant improvement of observational techniques in the past decades, more and more detailed
information about accretion disc systems could be gained.
In particular, a large number of ultraluminous X-ray sources (ULXs) have been discovered
(Fabbiano~1989), 
imposing a severe problem upon the existing general idea of accretion disc systems:
With a bolometric luminosity exceeding $10^{39}\,\textup{erg}\,\textup{s}^{-1}$,
at least some of them show relatively low radiation temperatures ($\sim 0.1\,\textup{keV}$).
These systems have been suggested to be intermediate mass black hole (IMBH),
sub-Eddington accretion disc systems (\cite{roberts_2005}).
However, large samples of ULXs 
from recent observational data reveal that a distinct class exists, showing higher temperatures --
sometimes exceeding $1\,\textup{keV}$ -- than can be explained by IMBHs (\cite{mizuno_1999}).
In contrast, stellar mass black holes accreting above their Eddington limit
can account for these sources (\cite{watarai_2001}).

In particular, the applicability of the classical Eddington limit
is an important point in this discussion:
Similar to the stellar case, disc accretion may be limited by
radiation pressure, counteracting gravity and viscous dissipation.
While in the stellar case, we are dealing with an approximately
spherically symmetric, i.\,e., 1-dimensional situation,
discs require an -- at least -- 2-dimensional treatment.
The stellar Eddington limit implies several assumptions,
which do not apply for the disc case, though it is not
clear {\it a priori\/} to what degree a proper treatment
will alter the resulting numbers.

In our work, we calculate the \emph{local} analog to the Eddington limit
for a classical thin disc model of the \cite{shakura_1973} variety and for a slim disc model of
the \cite{abramowicz_1988} type (Sect.~\ref{sec_bh_accretion}). In both descriptions, a standard
$\alpha$-viscosity is applied, therefore restricting the models to the
non-selfgravitating case. We also investigate the spectral
energy distribution of supercritical accretion flows in
the stellar mass black hole case, based on the radiation
hydrodynamic (RHD) simulations computed by \cite{ohsuga_2005} (Sect.~\ref{sec_sed}).
\section{Black hole accretion and the Eddington limit}\label{sec_bh_accretion}
The physical reasoning for the Eddington limit is an equilibrium of the
attractive force $F_\mytextup{g}$ and the repelling force
$F_\mytextup{r}$. The original stellar Eddington limit relies on
several assumptions: (1) spherical symmetry of the system;
(2) isotropic radiation; (3) homogeneous degree of
ionisation; (4) Thomson scattering as the sole source of
opacity; (5) negligible gas pressure; (6) no relativistic effects;
and (7) no time dependence (stationarity).
To investigate if these assumptions are an oversimplification
in the disc case, we relinquish the approximations (1)--(5),
but keep the approximation of non-relativistic stationarity
for our calculation.

We assume azimuthal symmetry in the following.
Therefore, one can define two different ``Eddington limits'',
corresponding to the equilibria in the vertical ($z$) and radial
($s$) direction.\footnote{We use a cylindrical coordinate system
$\{ s, \varphi, z \}$ with the distance $r$ to the origin, $r^2 = s^2 + z^2$}
For a rotating viscous disc, further contributions to the total
force have to be included in the radial direction:
\[
F^{(z)}_\mytextup{tot}(s) 
= F^{(z)}_\mytextup{r}(s) - F^{(z)}_\mytextup{g}(s) 
= 0\,,\qquad
F^{(s)}_\mytextup{tot}(s) 
= F^{(s)}_\mytextup{r}(s) - F^{(s)}_\mytextup{g}(s) + \ldots 
=0\,.
\]
In the following, as an example, we calculate the maximum amount of matter $\dot{M}_\mytextup{crit}$
that can be accreted towards the central object
in the case of a stellar mass black hole, $M = 10\,M_\odot$, for non-selfgravitating
discs with an $\alpha$-parameter of $0.1$. For both the thin and the slim disc case,
we use two descriptions for the opacity: Firstly, simple Thomson scattering
is applied: $\kappa_\mytextup{es} = 0.4\,\textup{cm}^2\textup{g}^{-1}$.
Secondly, we use an interpolation formula, $\kappa=\kappa_\mytextup{in}$ (Gail, priv. comm.),
which accounts for further contributions like bound-free and free-free absorption.
For a detailed description of $\kappa_\mytextup{in}$ see \cite{heinzeller_2006_2}.

We refer our results to the classical Eddington-limit, which in the disc case with a torque-free
boundary condition at the disc's inner radius $s_\mytextup{i} = 3 r_\mytextup{g} = 6GM/c^2$ is given by
$L_\mytextup{E} = 4 \upi c G M/\kappa_\mytextup{es}\allowbreak = 1.2 \cdot 10^{39}\,\textup{erg}/\textup{s}$,
and $\dot{M}_\mytextup{E} = 8 \upi c s_\mytextup{i}/\kappa_\mytextup{es}
= 2.6 \cdot 10^{-7}\,M_\odot/\textup{a}$.
Detailed calculations reveal that the vertical limit sets severer conditions
on the critical accretion rates and corresponding luminosities than the radial
limit in both the thin and the slim disc model. We therefore concentrate on
the results for the vertical limit in the following.

\begin{SCfigure}
\includegraphics[clip,width=0.45\textwidth]{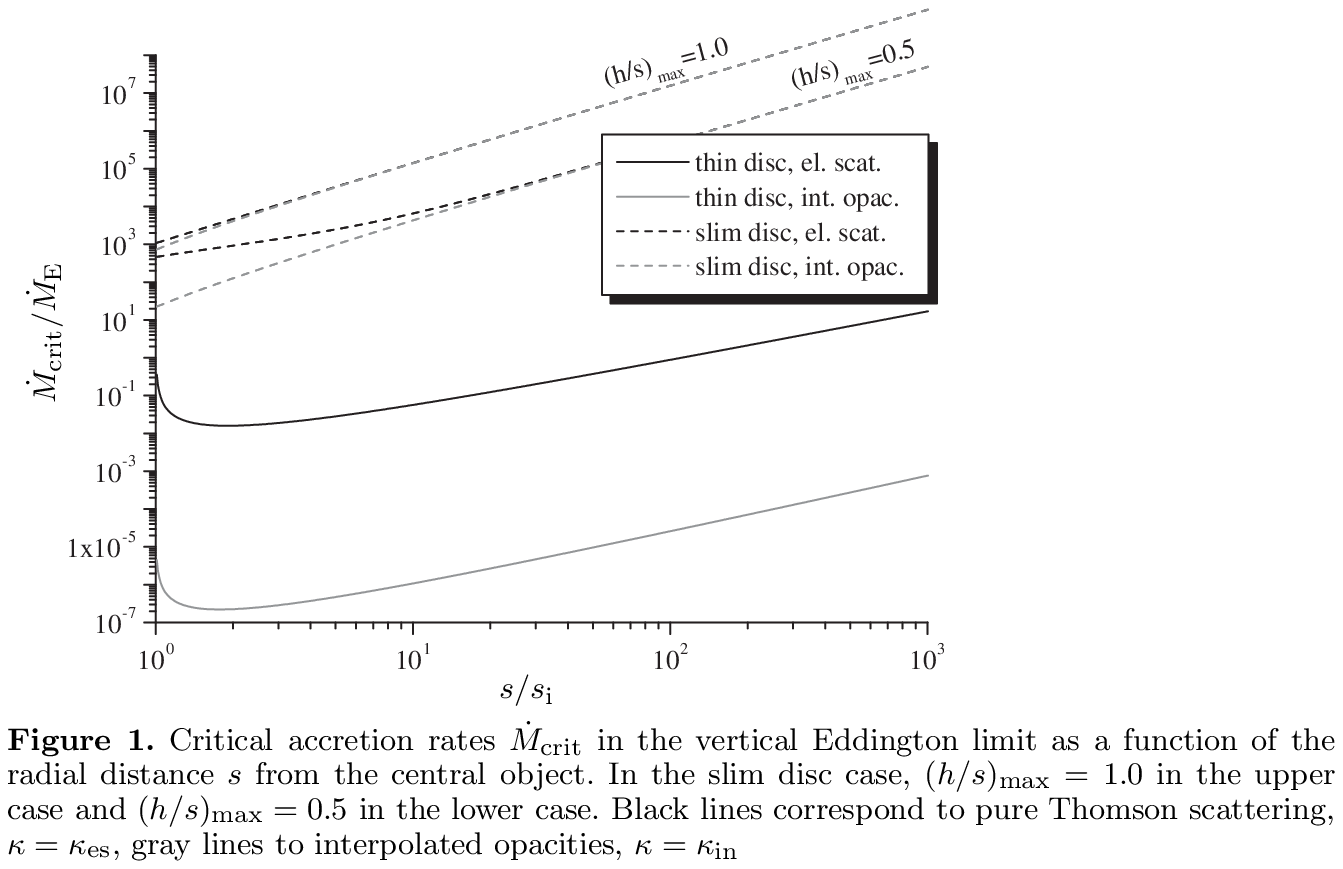}%
\caption{Critical local accretion rates in the vertical Eddington limit
as a function of the radial distance $s$ from the central object. For slim discs,
$(h/s)_\mytextup{max}=1.0$ in the upper and $(h/s)_\mytextup{max}=0.5$ in the lower case.
Black lines correspond to Thomson scattering, $\kappa=\kappa_\mytextup{es}$, gray lines
to interpolated opacities, $\kappa=\kappa_\mytextup{in}$}
\label{fig_mp_edd}
\end{SCfigure}
Figure~\ref{fig_mp_edd} shows the results for the critical accretion rates $\dot{M}_\mytextup{crit}$
in the vertical Eddington limit as a function of the radial distance $s$ from the central object.
In the slim disc case, one free parameter remains to be determined by hand: the ratio of the
local height $h$ of the disc to the local radial distance $s$. Since the slim disc model is valid
for $h/s \leq 1$, we plot the results for the two cases $(h/s)_\mytextup{max}=1.0$ and $(h/s)_\mytextup{max}=0.5$.

Obviously, the critical accretion rates are no longer given by a global
quantity like $\dot{M}_E$: they depend on the radius $\propto s^{1.2 \ldots 1.9}$. While the opacity has a strong influence on the
results ($\dot{M}_\mytextup{crit} \propto \kappa^{-1}$) for low accretion rates, it becomes
more and more unimportant in the slim disc model for high accretion rates. This is due to the
fact that advection dominates the energy transport in this regime. As can be seen from the
results for thin discs, the inner boundary condition plays a crucial role in determining the
maximum amount of matter which finally reaches the central black hole.
For slim discs, the critical accretion rates may reach up to $10^3 \dot{M}_\mytextup{E}$,
while at the same time the luminosity stays close to the classical limit,
$L \approx 20 L_E$ (\cite{watarai_2000}). Thus, highly super-Eddington accretion
comes along with only mildly super-Eddington luminosities.
\section{Spectral energy distribution of super-Eddington flows}\label{sec_sed}
\begin{figure}
~\includegraphics[width=0.45\textwidth,clip]{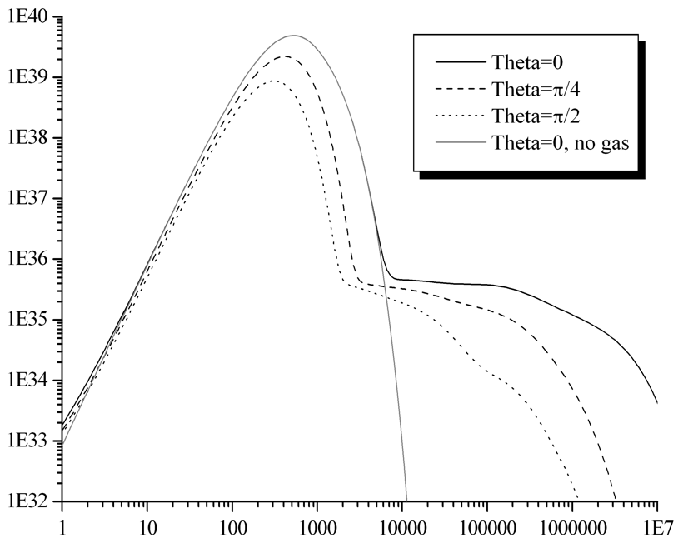}\hfill\includegraphics[width=0.45\textwidth]{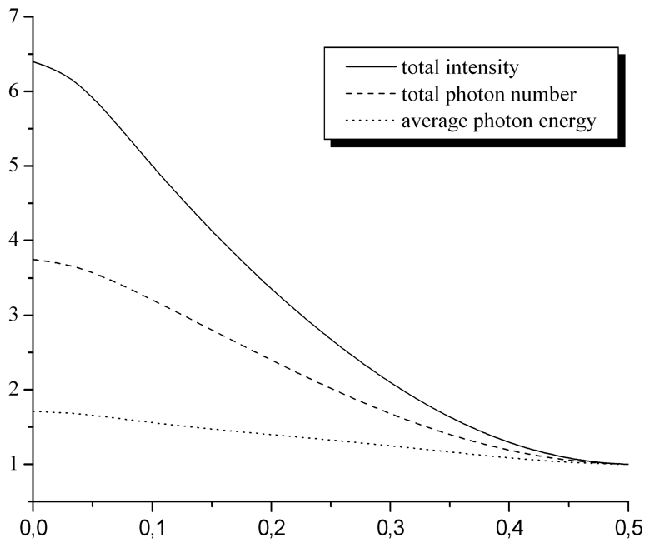}~%
\unitlength1mm%
\vskip-47mm%
\begin{picture}(100,50)(0,0)%
\put(0,2){\mbox{\footnotesize \textbf{(a)}}}%
\put(30,2){\mbox{\footnotesize $E_\gamma\,[\textup{eV}]$}}%
\put(0,23){\mbox{\footnotesize \begin{sideways} $\nu L_\nu\,[\textup{erg}/\textup{s}]$\end{sideways}}}%
\put(70,2){\mbox{\footnotesize \textbf{(b)}}}%
\put(102,2){\mbox{\footnotesize $\Theta/\upi$}}%
\end{picture}
\vskip-3mm
\caption{\textbf{(a)} Spectral energy distribution for various inclination angles $\Theta$; \textbf{(b)} Comparison of
total intensity, average photon energy and total photon number for $\Theta = 0 \ldots \upi/2$, normalized
to the corresponding values at $\upi/2$}\label{fig_sed}
\end{figure}
In the second part, we investigate the observational appearance of super-Eddington accretion discs.
This work is based on 2D RHD simulations, which are presented in detail in
\cite{ohsuga_2005}. In a spherical computational box of $500 R_\mytextup{G}$ radius,
matter is injected continuously in the disc plane $x-y$ around a $10 M_\odot$ black hole
at high rates ($\dot{M} = 10^3 \dot{M}_\mytextup{E}$). Again, the
viscosity is parameterized by $\alpha=0.1$. For the energy transport equation,
radiation and advection are taken into account, as well as photon
trapping effects (\cite{begelman_1978}).
We calculate the spectral energy distribution of this system for
various inclination angles $\Theta$ between $0$ (face-on) and $\upi/2$ (edge-on). 
For details about the computation, we refer the reader to~\cite{heinzeller_2006_1}.
The resulting SEDs for inclination angles $0$, $\upi/4$ and $\upi/2$ are shown in Fig.~\ref{fig_sed}(a).

While in the Rayleigh-Jeans part of the spectrum no difference between the orientations of the system
can be seen, the peak intensity region around $1\,\textup{keV}$ is shifted towards higher photon energies
and higher luminosities for lower inclination angles. For all orientations, a plateau-like structure
in the high-energy part of the spectrum can be observed, although this feature is more prominent for
the face-on case. Its origin lies in the emission of the hot gas in the vicinity
of the black hole: when calculating the spectrum for $\Theta=0$ while neglecting gas emission and absorption,
this plateau disappears.

To outline the orientation effects, we compare in Fig.~\ref{fig_sed}(b) the total intensity emitted by the system
for inclination angles $\Theta=\{0 \ldots \upi/2\}$. For small
inclinations, the intensity rises by a factor of up to $6.4$ for $\Theta=0$; this is
due to both an enhanced average photon energy and an increased total photon number. We identify this behavior with
\emph{mild relativistic beaming}. When fitting the peak intensity region with a black body spectrum,
we derive temperatures between  $1.1 \cdot 10^6\,\textup{K}$ for $\Theta=\upi/2$
and $1.7 \cdot 10^6\,\textup{K}$ for $\Theta=0$.
\section{Conclusions}
The above investigations allow us to draw several conclusions on black hole accretion processes.
Firstly, we show that the classical Eddington limit can not be applied in discs. Contrary,
the critical accretion rates become a local quantity, depending on the distance from the central object.
Thereby, the inner disc region and inner boundary hold the key position in determining
the final amount of matter that can be accreted by the central object. We find that super-Eddington
accretion is possible with accretion rates up to $10^3 \dot{M}_\mytextup{E}$, corresponding to slightly
super-Eddington luminosities up to $20 L_\mytextup{E}$. Since the bottle-neck for accretion lies
in the inner region of the disc, our results may also correlate to the question on the origin of outflows and
jets. Secondly, our work on the spectral energy distribution of such super-Eddington accreting systems shows
that for a proper interpretation of the spectrum, the combined system of the disc and its surroundings
has to be taken into account. Features like the high-energy plateau can not be reproduced by considering
the accretion disc only. Also, we find that mild relativistic beaming becomes important for small inclination
angles, leading to higher luminosities and also higher temperatures. However, these temperatures are still
too low to account for the observed ULX sources which sometimes reach temperatures up to $10^7\,\textup{K}$.
This may be due to the lack of Comptonization effects in our calculations:
\cite{socrates_2004}, for example, showed that turbulent Comptonization produces a significant
contribution to the far-UV and X-ray emission of black hole accretion discs. Thus, the inclusion of
Comptonization effects will be a primary goal in our future investigations.


\begin{thebibliography}{}
\bibitem[Abramowicz, Czerny, Lasota, \etal\ (1988)]{abramowicz_1988}
{Abramowicz, M.A., Czerny, B., Lasota, J.P., Szuzkiewicz, E.} 1988, \textit{ApJ} {332}, 646
\bibitem[Begelman (1978)]{begelman_1978}
{Begelman, M.C.} 1978, \textit{MNRAS} 184, 53
\bibitem[Fabbiano (1989)]{fabbiano_1989}
{Fabbiano, G.} 1989, \textit{ARA\&A} 27, 87
\bibitem[Heinzeller \& Duschl (2006)]{heinzeller_2006_2}
{Heinzeller, D. \& Duschl, W.J.} 2006, to appear in \textit{MNRAS}
\bibitem[Heinzeller, Mineshige \& Ohsuga (2006)]{heinzeller_2006_1}
{Heinzeller, D., Mineshige, S. \& Ohsuga, K.} 2006, to appear in \textit{MNRAS}
\bibitem[Mizuno, Ohnishi, Kubota, \etal\ (1999)]{mizuno_1999}
{Mizuno, T., Ohnishi, T., Kubota, A., Makishima, K., Tashiro, M.} 1999, \textit{PASJ} 51, 663
\bibitem[Ohsuga, Mori, Nakamoto, \etal\ (2005)]{ohsuga_2005}
{Ohsuga, K., Mori, M., Nakamoto, T., Mineshige, S.} 2005, \textit{ApJ} 628, 368
\bibitem[Roberts, Warwick, Ward, \etal\ (2005)]{roberts_2005}
{Roberts, T.P., Warwick, R.S., Ward, M.J., Goad, M.R., Jenkins, L.P.} 2005, \textit{MNRAS} 357, 1363
\bibitem[Shakura \& Sunyaev (1973)]{shakura_1973}
{Shakura, N.I. \& Sunyaev, R.A.} 1973, \textit{A\&A} 24, 337
\bibitem[Socrates, Davis \& Blaes (2004)]{socrates_2004}
{Socrates, A., Davis, S.W. \& Blaes, O.} 2004, \textit{ApJ} 601, 405
\bibitem[Watarai, Fukue, Takeuchi, \etal\ (2000)]{watarai_2000}
{Watarai, K.-y., Fukue, J., Takeuchi, M., Mineshige, S.} 2000, \textit{PASJ} 52, 133
\bibitem[Watarai, Mizuno \& Mineshige (2001)]{watarai_2001}
{Watarai, K.-y., Mizuno, T. \& Mineshige, S.} 2001, \textit{ApJ} 549, L77
\end{thebibliography}
\end{document}